\documentclass[10pt,twocolumn,letterpaper]{article}

\usepackage[pagenumbers]{cvpr} 

\usepackage[dvipsnames]{xcolor}

\definecolor{cvprblue}{rgb}{0.21,0.49,0.74}
\usepackage{graphicx}
\usepackage{nicematrix,enumitem}
\usepackage{booktabs}
\usepackage{multirow}
\usepackage[tableposition=top]{caption}
\usepackage{subcaption}
\usepackage[pagebackref,breaklinks,colorlinks,citecolor=cvprblue]{hyperref}

\def\paperID{38} 
\def\confName{CVPR}
\def\confYear{2024}

\title{Improving Breast Cancer Grade Prediction with Multiparametric MRI Created Using Optimized Synthetic Correlated Diffusion Imaging}

\author{Chi-en Amy Tai \\
University of Waterloo\\
Waterloo, ON\\
{\tt\small amy.tai@uwaterloo.ca}
\and Alexander Wong \\
University of Waterloo\\
Waterloo, ON\\
{\tt\small alexander.wong@uwaterloo.ca}
}

\begin{document}
\maketitle
\begin{abstract}
Breast cancer was diagnosed for over 7.8 million women between 2015 to 2020. Grading plays a vital role in breast cancer treatment planning. However, the current tumor grading method involves extracting tissue from patients, leading to stress, discomfort, and high medical costs. A recent paper leveraging volumetric deep radiomic features from synthetic correlated diffusion imaging (CDI\textsuperscript{s}) for breast cancer grade prediction showed immense promise for noninvasive methods for grading. Motivated by the impact of CDI\textsuperscript{s} optimization for prostate cancer delineation, this paper examines using optimized CDI\textsuperscript{s} to improve breast cancer grade prediction. We fuse the optimized CDI\textsuperscript{s} signal with diffusion-weighted imaging (DWI) to create a multiparametric MRI for each patient. Using a larger patient cohort and training across all the layers of a pretrained MONAI model, we achieve a leave-one-out cross-validation accuracy of 95.79\%, over 8\% higher compared to that previously reported. 
\end{abstract}    
\section{Introduction}
\label{sec:intro}

Breast cancer was diagnosed for over 7.8 million women between 2015 to 2020 and resulted in 685,000 deaths across the world in 2020~\cite{bca-who-stats}. Grading is a crucial factor in breast cancer treatment planning, but the current method to grade breast cancer tumors involves tissue extraction from the patient which causes patient stress and discomfort~\cite{biospy-pain}, along with high medical fees~\cite{breast-biopsy-costs}. In addition, the current method to determine the cancer grade is based on a human pathologist's opinion~\cite{bc-grading}. With potential biases and high uncertainty in clinical judgement, it is possible for the patient to receive an incorrect grade leading to an unsuitable treatment strategy~\cite{human-judgement-problem}. Recent advancements in computer vision and imaging in the cancer domain have shown promise of noninvasive ways to diagnose and evaluate cancer tumours with high accuracy~\cite{tai2023enhancing}.

In literature, multiparametric MRIs have recently gained popularity as they have been shown to greatly benefit breast cancer clinical task enhancement with deep learning~\cite{marino2018multiparametric}. Multiparametric MRIs are created by fusing two or more MRI modalities together with the combination achieving better performance than each MRI modality on its own~\cite{hu2020deep}. 

Specifically, leveraging volumetric deep radiomic features from synthetic correlated diffusion imaging (CDI\textsuperscript{s}) fused with DWI for breast cancer grade prediction showed promise compared to other MRI modalities, achieving a prediction accuracy of 87.70\%~\cite{tai2023enhancing}. First introduced in Wong et al.~\cite{wong2022synthetic}, CDI\textsuperscript{s} is a method that utilizes a combination of native and synthetic diffusion signal acquisitions, coupled with a signal calibration that aims to achieve enhanced consistency in the dynamic range across various machines and protocols. CDI\textsuperscript{s}, on its own, was shown to produce superior results for prostate cancer delineation compared to other gold-standard modalities. 

In Wong et al.~\cite{wong2022synthetic}, they demonstrate the impact of tuning CDI\textsuperscript{s} for the specific cancer domain. However, in Tai et al.~\cite{tai2023enhancing}, they only slightly modified CDI\textsuperscript{s} for breast cancer instead of properly tuning the coefficients. Motivated by the impact of CDI\textsuperscript{s} optimization for prostate cancer delineation in~\cite{wong2022synthetic}, this paper examines using optimized CDI\textsuperscript{s} to improve breast cancer grade prediction.

\section{Methodology}
\label{sec:methodology}
\begin{table}
    \caption{SBR grade distribution in the patient cohort.}
    \centering
    \begin{NiceTabular}{l c}
    \toprule
    \RowStyle{\bfseries}
    \textbf{SBR Grade} & \textbf{Number of Patients} \\ \midrule
    Grade I (Low) & 10 \\
    Grade II (Intermediate) & 99 \\
    Grade III (High) & 200 \\
    \bottomrule
    \end{NiceTabular}
 \label{tab:sbr-grade-dist-v2}
\end{table}

\begin{figure*}
    \centering
    \includegraphics[width=\linewidth]{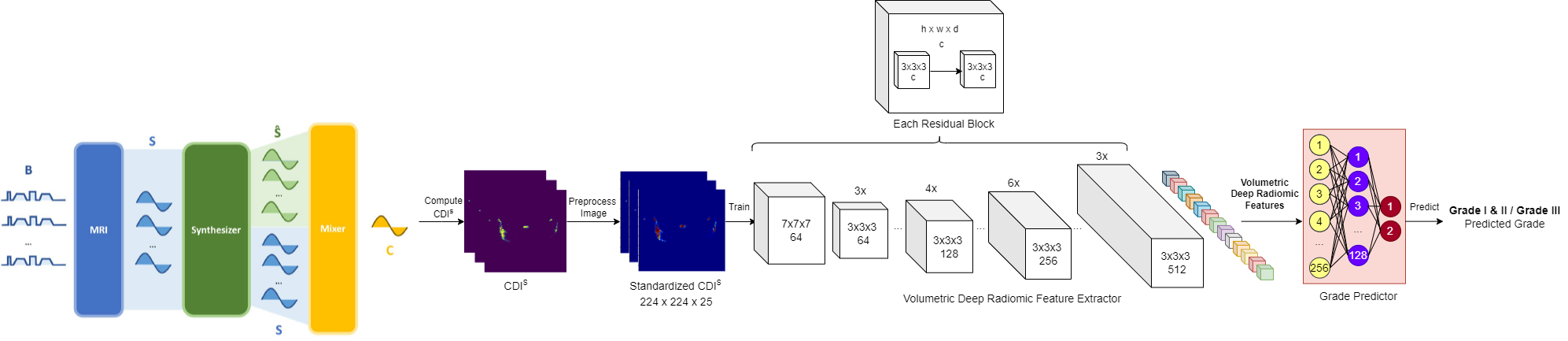}
    \caption{Clinical support workflow copied from~\cite{tai2023enhancing} that is used in this study.}
    \label{fig:support-workflow}
\end{figure*}

In this study, we use the American College of Radiology Imaging Network (ACRIN) 6698/I-SPY2 study~\cite{acrin6698-data-1, acrin6698-data-2, acrin6698-data-3, acrin6698-data-4} and filter for non-null Scarff-Bloom-Richardson (SBR) grade values with a total of 309 patients remaining. Similar to~\cite{tai2023enhancing}, we also combine SBR grade I and II into one category due to the imbalanced distribution between the three grades (as shown in Table~\ref{tab:sbr-grade-dist-v2}). 

To obtain the optimized CDI\textsuperscript{s} coefficients, we leverage the Nelder-Mead simplex optimization strategy to maximize the area under the receiver operating characteristic curve (AUC) for breast cancer tumour delineation. After generating the optimized CDI\textsuperscript{s} signals, we fuse the signals on the diffusion-weighted imaging (DWI) to create a multiparametric MRI. To achieve dimensional consistency for machine learning, all volumes are then standardized into 224x224x25 volumetric data cubes for each patient.

Using a previously introduced deep radiomic clinical support workflow~\cite{tai2023enhancing} (shown in Figure~\ref{fig:support-workflow}), we leverage a pretrained 34-layer volumetric residual convolutional neural network architecture initialized with weights from MONAI~\cite{monai} and train it to extract deep radiomic features. The MONAI weights were derived by training on the extensive 3D medical dataset, 3DSeg-8. This comprehensive dataset comprises images from eight different 3D segmentation datasets, encompassing both MRI and CT images~\cite{chen2019med3d}. These features are then fed into a fully-connected neural network grade predictor to predict breast cancer grade (Grade I/Grade II and Grade III). 

For training, a learning rate of 0.001 was used along with a weighted random sampler, AdamW optimizer, cross-entropy loss function and cosine annealing learning rate scheduler. We also train all the model layers with no freezing. Leave-one-out cross-validation was conducted to obtain the results with the average accuracy, sensitivity, and specificity recorded. 

\section{Results}
\label{sec:results}
\begin{figure}
    \centering
    \subfloat[]{\includegraphics[width=0.48\linewidth]{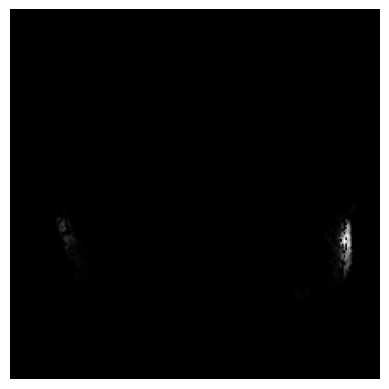}} 
    \subfloat[]{\includegraphics[width=0.48\linewidth]{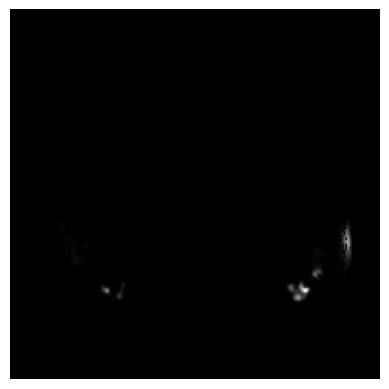}} \\
    \subfloat[]{\includegraphics[width=0.48\linewidth]{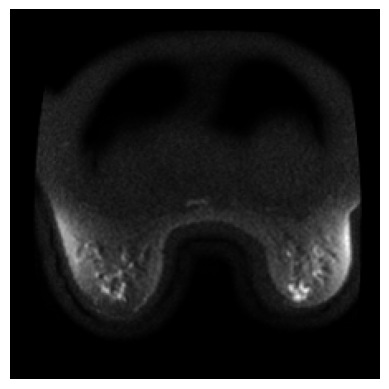}} 
    \subfloat[]{\includegraphics[width=0.48\linewidth]{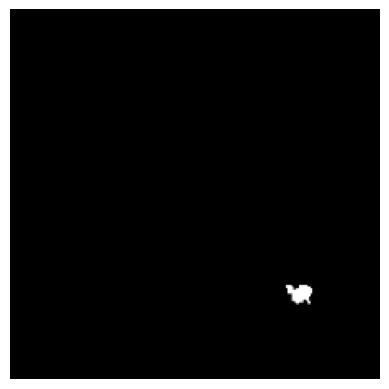}} 
    \caption{An example slice illustrating visual differences between (a) Unoptimized CDI\textsuperscript{s}, (b) Optimized CDI\textsuperscript{s}, (c) the associated DWI, and (d) the associated tumour mask for a patient who has SBR Grade III (High). In this patient case, grade prediction was correct using the Optimized CDI\textsuperscript{s} signal fused with DWI.}
    \label{fig:v2-grade-comparison-results}
\end{figure}

As seen in Table~\ref{tab:sbr-modalities-monai-resnet-v2}, using the optimized CDI\textsuperscript{s} obtained a leave-one-out cross-validation of 95.79\%, over 8\% higher than that previously reported (87.70\%). Optimizing CDI\textsuperscript{s} also achieved higher sensitivity and specificity, with all values above 90\%. Figure~\ref{fig:v2-grade-comparison-results} shows an illustrative example highlighting the visual differences between an unoptimized CDI\textsuperscript{s} signal and an optimized CDI\textsuperscript{s} signal, along with the corresponding tumour mask and DWI. Given the promising results, this proposed noninvasive method to identify the severity of cancer would allow for better treatment decisions without the need for a biopsy and highlights the impact of optimizing CDI\textsuperscript{s} for the specific cancer domain. 

\begin{table}
    \caption{Results using unoptimized and optimized CDI\textsuperscript{s} with the best results bolded.}
    \centering
    \begin{NiceTabular}{l c c c}
        \toprule
        \RowStyle{\bfseries}
        CDI\textsuperscript{s} Version & Accuracy & Sensitivity & Specificity \\ \midrule
        Unoptimized~\cite{tai2023enhancing} & 87.70\% & 90.29\% & 81.82\% \\
        Optimized & \textbf{95.79\%} & \textbf{96.50\%} & \textbf{94.50\%} \\
        \bottomrule
    \end{NiceTabular}
    \label{tab:sbr-modalities-monai-resnet-v2}
\end{table}

{
    \small
    \bibliographystyle{ieeenat_fullname}
    \bibliography{main}
}

\end{document}